\documentclass[final]{aipproc}

\layoutstyle{6x9}

\begin{document}

\title{Alpha-Alpha scattering, chiral symmetry and $^8$Be lifetime\footnote{Talk presented at
{\em SCADRON 70 Workshop on ``Scalar Mesons and Related Topics''},
Lisbon, 11-16 February 2008}}


\author{E. Ruiz Arriola}{
  address={Departamento de F\'{\i}sica At\'omica, Molecular y
Nuclear \\ Universidad de Granada \\ E-18071 Granada, Spain.}
}

\begin{abstract}
Alpha-alpha scattering is discussed in terms of a chiral two pion
exchange potential (TPE) which turns out to be attractive and singular
at the origin, hence demanding renormalization. When $^8 {\rm Be}$ is
treated as a resonance state a model independent correlation between
the Q-factor and lifetime $1/\Gamma$ for the decay into two alpha
particles arises. For a wide range of parameters compatible with
potential model analyses of low energy $\pi \alpha $ scattering it is
found $\Gamma = 4.4(4) {\rm eV}$ in fairly good agreement with the
experimental value $\Gamma_{\rm exp.} = 5.57 (25) {\rm eV}$. The
remaining discrepancy as well as the phase shift up to $E_{\rm LAB}=15
{\rm MeV}$ could be accommodated by the leading nuclear peripheral
contributions due to the $^3$H+p and $^3$He+n continuum.
\end{abstract}

\maketitle

\begin{figure}
\includegraphics[height=.25\textheight]{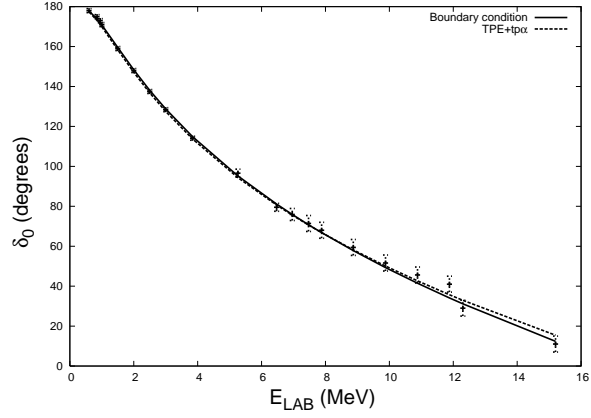}
\caption{s-wave $\alpha\alpha$ phase-shifts. We show the boundary
condition (solid line), and the TPE+$t p\alpha$+$^3{\rm He} n\alpha$
corrections (dashed line). Data are from~\cite{RevModPhys.41.247}.}
\label{fig:phaseshift}
\end{figure}

{\bf 1.} The scattering of $\alpha$-particles has been one of the most
studied nuclear reactions both theoretically and
experimentally~\cite{RevModPhys.41.247} and reveals the existence of
$^8{\rm Be}$ as a narrow $ (J^P, T)= (0^+,0)$ threshold resonance at
CM energy $Q=91.84(4) {\rm KeV}$ and a very small (Breit-Wigner) width
$\Gamma_{BW}=5.57(25) {\rm eV}$~(see \cite{Tilley:2004} for a review)
which corresponds to a life-time of $\tau= 1.18 \time 10^{-16} s $.
The $^8$Be state is so close to the $\alpha\alpha$-threshold that the
scattering length is $\alpha_0=-1600 {\rm fm} $, a huge scale in
nuclear physics. This requires a great deal of fine tuning in the
interaction, an issue relevant for the invokers of the Anthropic
Principle.  At the resonant energy the corresponding de Broglie
wavelength is $1/p = 1/\sqrt{M_\alpha Q } \sim 10 {\rm fm}
$~\footnote{We take $M_\alpha=2(M_p+M_n) - B = 3727.37 {\rm MeV} $,
corresponding to a binding energy $B=28.2957 {\rm MeV}$.  } much
larger than the size of the $\alpha$-particle so one would not expect
internal structure playing a crucial role.  We illustrate this for the
s-wave state below the inelastic $^7$Li+p and $^7$B+n thresholds which
satisfies
\begin{eqnarray}
-u_{p}''(r) &+& \left[ M_\alpha V_{\alpha \alpha} (r) + \frac{2}{a_B
 r} \right] u_{p}(r) = p^2 u_{p} (r) \, , 
\label{eq:sch-l}
\end{eqnarray} 
with $a_B = 2/ (M_\alpha Z_\alpha^2 e^2) = 3.63 {\rm fm}$ the Bohr
radius and $p = \sqrt{ M_\alpha E} $ the CM momentum.  If we 
implement the strong interaction by an energy independent short
distance boundary condition a fit of the s-wave scattering phase
shifts~\cite{RevModPhys.41.247} up to $E_{\rm LAB} = 15 {\rm MeV}$
becomes possible (see Fig.~\ref{fig:phaseshift}) yielding the result 
\begin{eqnarray}
u_p'(r_c) /u_p (r_c) = - 0.357(3) {\rm fm}^{-1} \qquad r_c = 2.88(3) {\rm fm} \qquad \chi^2 /DOF = 0.5
\label{bc:fit} 
\end{eqnarray} 
The value of the Coulomb potential at $r_c=3 {\rm fm}$ is $V_C (r_c)
\sim 1 {\rm MeV}$. Actually, if we switch it off down to $r_c$ we get
a would-be $^8$Be bound state, $u_B(r) =A_B e^{-\gamma_B r}$ where
$B=-\gamma_B^2 / M_\alpha = 1.3 {\rm MeV}$. So, the nuclear attraction
must cancel the $\sim 1 {\rm MeV}$ electromagnetic energy to allow the
$\alpha\alpha$ system to tunnel through the barrier and eventually
form $^8$Be. The $\alpha$-particle m.s.r. as measured in electron
scattering~\cite{PhysRev.160.874} is $ r_\alpha^{\rm em } = 1.668(5)
{\rm fm}$ and due to isospin symmetry the matter density is $\rho_B
(r) = 2 \rho_{\rm em } (r) $, so at relative distances of $r\sim {\rm
3 fm}$ we do not expect finite size effects to be crucial, as the
boundary condition model fit, Eq.(\ref{bc:fit}), explicitly shows.

{\bf 2.} Besides the Coulomb interaction it seems unclear what does
effectively make two alpha particles attract each other. In addition
to the simple boundary condition described above there are various
potentials describing the data successfully up to similar or higher
energies; an attractive square well potential plus hard
core~\cite{Kermode:1965}, a Woods-Saxon potential~\cite{Kukulin:1971}
as well as a Gaussian potential (see e.g.\cite{Friedrich:1981ad} for a
review on microscopic approaches). Here we analyze {\it model
independent} attraction mechanisms based on the assumption of a
marginal influence of internal $\alpha$-particle structure.

\begin{figure}
\includegraphics[height=.15\textheight]{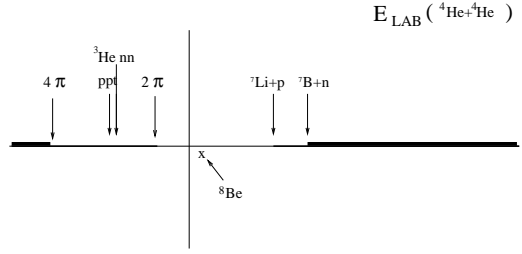}
\caption{Singularity exchange left cuts and unitarity right cuts for the 
$\alpha\alpha$-system as a function of the LAB energy. The $^8$Be
resonance corresponds to a pole on the Second Riemann sheet.}
\label{fig:complex}
\end{figure}

The longest possible particle exchange corresponds to two pions (one
pion exchange is forbidden because of parity and isospin), which
yields an exponential tail $\sim e^{- 2m_\pi r}$ with the scale $1/(2
m_\pi)= 0.7 {\rm fm}$. This looks like too short as compared to the
$\alpha$-particle size, but note that it is only necessary that such a
contribution cancels the Coulomb barrier when the two alphas are
nearly touching or slightly overlapping. Obviously we do not expect to
{\it see} TPE inside the nucleus. A shorter range scale is provided by
the $t+p$ and $^3$He+n continuum. These and other scales generate left cuts in
the $\alpha\alpha$ scattering partial wave amplitudes
for the complex LAB energy plane as illustrated in
Fig.~\ref{fig:complex} where also the right cuts due to openning of
$^7$Li+n and $^7$B+n channels are displayed.

{\bf 3.} The idea of associating elementary fields to nuclei at low
energies complies to a manifest irrelevance of internal
structure~\cite{Locher:1978dk}. The role of chiral symmetry, essential
to describe long range pion exchanges, has been addressed only
recently~\cite{Arriola:2007de}. We take a scalar-isoscalar
Klein-Gordon charged as well as chirally invariant under $SU(2)_R
\otimes SU(2)_L $ transformations field for the $^4{\rm He}$
nucleus. The effective Lagrangean will include
pions~\cite{Weinberg:1978kz} and $\alpha$ particles which being much
heavier, $M_\alpha \gg m_\pi$, are better treated by transforming the
Klein-Gordon field as $\alpha (x) = e^{-i M_\alpha v \cdot x} \alpha_v
(x) $ with $\alpha_v (x) $ the heavy field and $v^\mu$ a four-vector
fulfilling $v^2=1$, eliminating the heavy mass
term~\cite{Jenkins:1990jv}. Keeping the leading $M_\alpha$ term, the
effective Lagrangean reads
\begin{eqnarray}
{\cal L} &=& 2 i M_\alpha \bar \alpha_v v \cdot \partial \alpha_v +
\frac{f^2}{4} \left[ \langle \partial^\mu U^\dagger \partial_\mu U
\rangle + \langle \chi U^\dagger + \chi^\dagger U \rangle \right]
\nonumber \\ &+& g_0 \bar \alpha_v \alpha_v \langle \partial^\mu
U^\dagger \partial_\mu U \rangle + g_1 \bar \alpha_v \alpha_v \langle
\chi U^\dagger + \chi^\dagger U \rangle + g_2 \bar
\alpha_v \alpha_v \langle v \cdot \partial U^\dagger v \cdot \partial
U \rangle 
\label{eq:L_I} 
\end{eqnarray} 
where the pion field in the non-linear representation is written as a
$SU(2)$-matrix, $U= e^{ i \vec \tau \cdot \vec \pi / f_\pi} $, with
$\vec \tau$ the isospin Pauli matrices, $f_\pi$ the pion weak decay
constant $f_\pi = 92.6 {\rm MeV}$, $\chi= m_\pi^2 /2 $ and $\langle ,
\rangle $ means trace in isospin space.  Here $g_0$, $g_1$, $g_2 $ are
dimensionless coupling constants which are not fixed by chiral
symmetry. This Lagrangean is the analog of the Weinberg-Tomozawa
Lagrangean and EFT extensions for $\pi N $
interactions~\cite{Ericson:1988gk} to the case of the $\pi \alpha $
system.  Photons are included by standard minimal coupling
$\partial^\mu \alpha \to D^\mu \alpha = \partial^\mu \alpha + Z_\alpha
e i A^\mu \alpha $.

\begin{figure}
\includegraphics[height=.15\textheight]{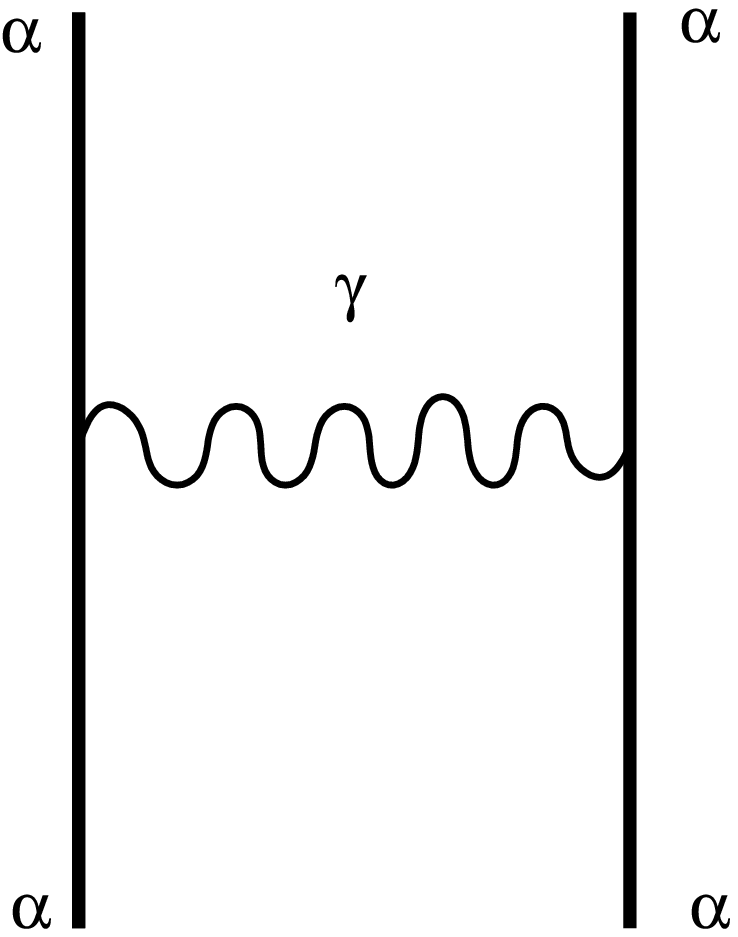}
\includegraphics[height=.15\textheight]{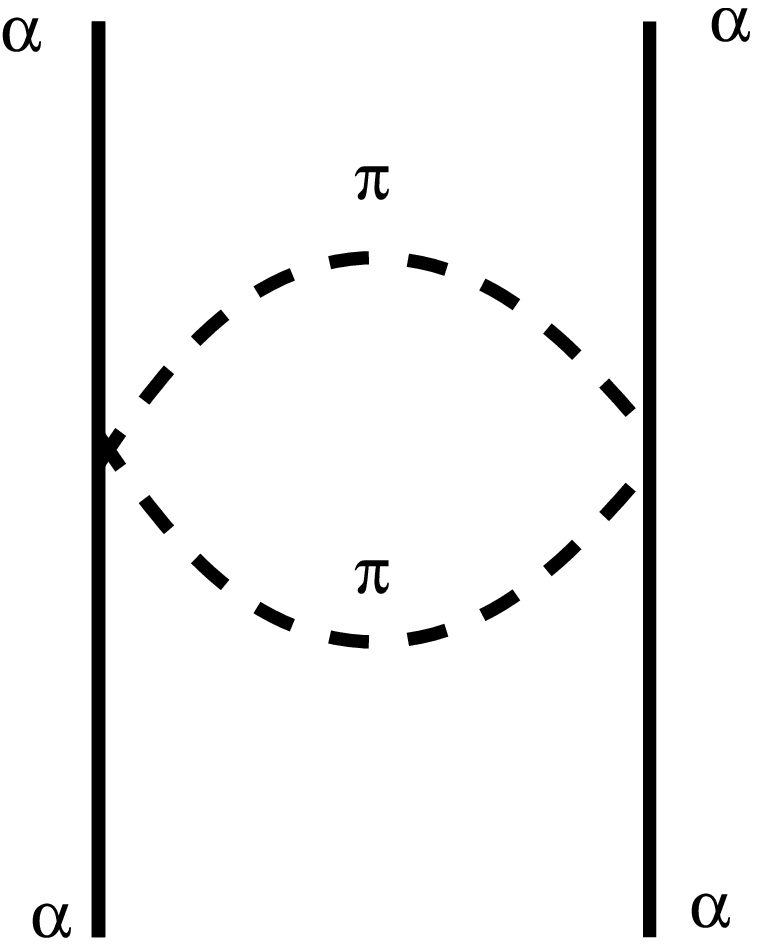}
\includegraphics[height=.15\textheight]{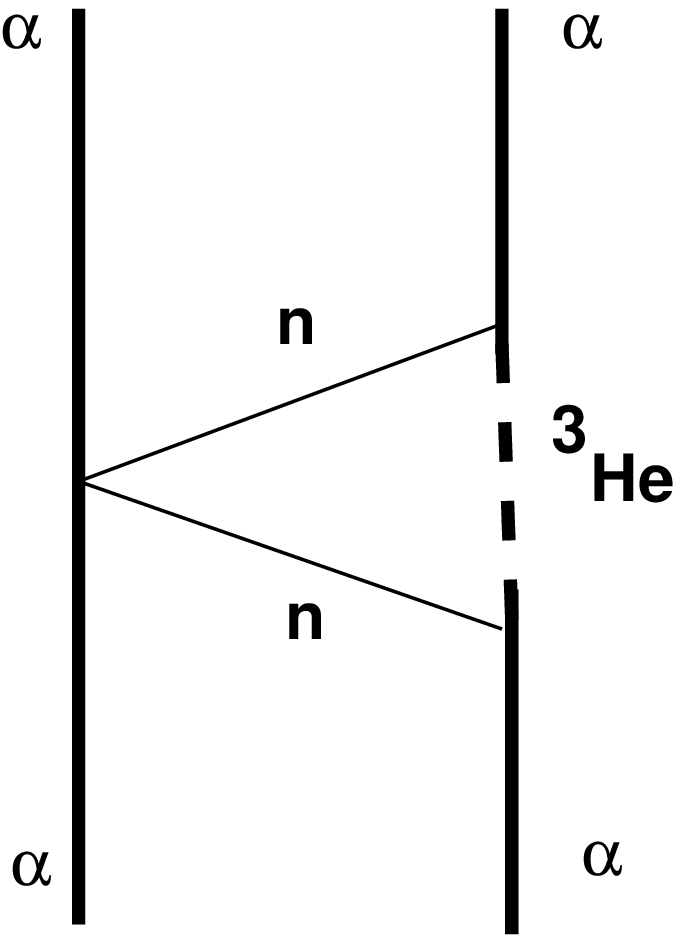}
\includegraphics[height=.15\textheight]{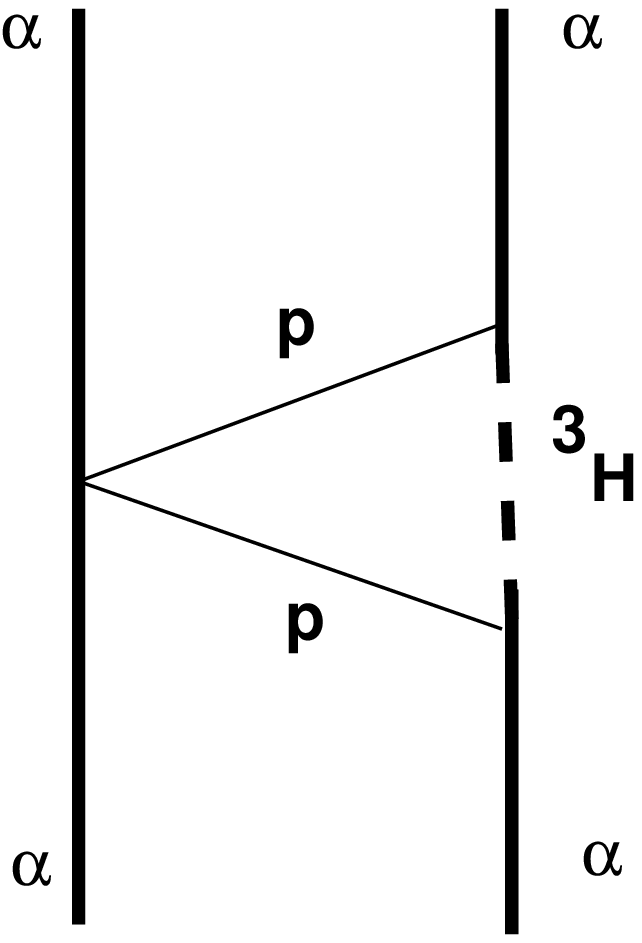}
\caption{Diagrams contributing to the $\alpha-\alpha$ potential at
 long distances (from left to right) : One-Photon Exchange, Two-Pion
 Exchange loop, $p$ and $n$ exchange loops.}
\label{fig:tpe}
\end{figure}


{\bf 4.} The leading direct t-channel TPE contribution to the scattering
amplitude is depicted in Fig.~\ref{fig:tpe}. The final result
reads~\cite{Arriola:2007de}
\begin{eqnarray}
V_{\alpha \alpha}^{2\pi} (r)&=&-\frac{3 m_\pi^7 \left[ K_0 ( 2 x) f(x) + K_1(2 x) g(x)  \right] }{ 32 \pi^3
M_\alpha^2 f_\pi^4 x^6}  
\label{eq:tpe-pot}
\end{eqnarray}
where $x= m_\pi r$, $K_0 (x)$ and $K_1(x)$ are modified Bessel
functions and $ f(x) = 4(g_0+g_1)^2 x^4 + 10( 12 g_0^2 + 4g_2 g_0 + 3
g_2^2) + \left[84 g_0^2 + 24 (g_1+g_2) g_0 + g_2 (4 g_1 +15 g_2)
\right] x^2 $, $ g(x) = 4 (g_0 + g_1)( 6 g_0 + g_2) x^3 + 10(12 g_0^2
+ 4 g_0 g_2 + 3 g_2^2)x $. The TPE potential is attractive
everywhere. In fact, the TPE potential becomes singular at short
distances,
\begin{eqnarray}
V_{\alpha \alpha}^{2\pi} (\vec x ) = - \frac{15 ( 12 g_0^2 + 4 g_0 g_2
+ 3 g_2^2)}{32 \pi^3 M_\alpha^2 f_\pi^4 } \, \frac{1}{r^7} + \dots
\label{eq:v-short}
\end{eqnarray} 
This is a relativistic and attractive Van der Waals interaction which
is explicitly independent on the pion mass. 
In the opposite limit of long distances we have
\begin{eqnarray}
V_{\alpha \alpha}^{2\pi} (\vec x ) \to - \frac{3 (g_0+g_1)^2
m_\pi^{9/2}}{16 \pi^{5/2} M_\alpha^2 f_\pi^4 } \frac{e^{-2 m_\pi 
r}}{r^{5/2}} \, . 
\label{eq:v-long}
\end{eqnarray}
The couplings $g_0$, $g_1$ and $g_2$ may be estimated from the
analysis of low energy $\pi\alpha$ scattering after pion-absorption and
Coulomb effects are eliminated~\cite{Khankhasaev:1989xf} yielding $g_0
= -82(11)$, $g_1=-5.3(3)$ and $g_2=77(12)$. Pion exchange interactions
between $\alpha$ particles have been treated in the past in a variety
of ways. A resonating group method approach was used in
Ref.~\cite{Shimodaya:1960} with an approximation for the TPE in the
mid-range. Forward dispersion relations for $\alpha\alpha$ scattering
have been discussed~\cite{FangLandau:1974wh} with the TPE cut replaced
by a pole. A folding model from a NN potential was used in
Ref.~\cite{Ericson:1981sa} and the $I$-wave phase shift was computed
in first order perturbation theory. This corresponds to our 
potential but missing the term with $g_2$.


{\bf 5.} Singular potentials are commonplace in effective theories.
Fortunately, one may use renormalization to actually {\it minimize}
our lack of knowledge at short distances, which we expect to be
irrelevant. This viewpoint in fact refuses to explain the fine
tunings, but makes use of their existence to insure short distance
insensitivity given some reliable long distance physics. In our case,
the problem is to solve Eq.~(\ref{eq:sch-l}) with suitable boundary
conditions. For the singular power-like potential~(\ref{eq:v-short}) written
as $M_\alpha V (r) \sim -R^5/r^7 $ ($R \sim 1.5 {\rm fm}$) one can
show that at short distances the wave function goes as
\begin{eqnarray}
u_{p} (r) &\to & C \left(\frac{r}{R}\right)^{7/4} \sin\left[
\frac{2}{5} \left(\frac{R}{r}\right)^{\frac{5}2} + \varphi \right] 
\, .
\label{eq:phase}
\end{eqnarray}
where the arbitrary phase $\varphi$ must be real and energy
independent.  As we see, the regularity condition, $u_{p}(0)=0$,
does not fix the solution uniquely~\cite{Case:1950,Frank:1971xx}, thus
some information must be provided in addition to the potential. 
For instance, if we choose the $Q$ we can determine $\Gamma$ and the
full phase shift.  This is the non-perturbative
renormalization program with one counterterm described
in~\cite{Valderrama:2005wv} for singular potentials within the NN
context. In practice, it is more convenient to introduce a short
distance cut-off $r_c \ll 1 {\rm fm}$ and to use a {\it real} and {\it
energy independent} boundary condition
\begin{eqnarray}
{\rm Re} \left[ \frac{u_{p}'(r_c)}{u_{p}
(r_c)}\right] = \frac{u_{p}'(r_c)}{u_{p}
(r_c)} = \frac{u_{0}'(r_c)}{u_{0}
(r_c)} \, . 
\label{eq:log-corr} 
\end{eqnarray} 
which implements self-adjointness in the domain $r \ge r_c$.

{\bf 6.} The experimentally determined~\cite{Tilley:2004} Breit-Wigner
small width involves the s-wave phase
shift~\cite{Rasche:1967,Kermode:1967} $\Gamma_{\rm BW} = 2 / \delta'_0
(E_R)$ for $\delta_0 ( E_R ) =\pi /2$. We get
\begin{eqnarray}
\Gamma_{\rm BW} ( ^8 {\rm Be} \to \alpha \alpha)= 4.3(3) \, {\rm eV}
\, , \, (\, {\rm exp.} 5.57 (25) {\rm eV}\, ) \, ,
\end{eqnarray} 
for $E_R =91.8 {\rm KeV}$ and the couplings $g_0$, $g_1$ and $g_2$
with their uncertainties obtained from low energy $\pi \alpha$
scattering~\footnote{A WKB analysis was also
done~\cite{Arriola:2007de} yielding $\Gamma_{\rm WKB} = 8.6 (4) {\rm
eV}$ for the experimental $Q$}.

Nevertheless, a rigorous treatment of $^8 {\rm Be} $ as a
exponentially time-decaying  state requires finding a pole of the
S-matrix in the second Riemann sheet of the complex energy plane,
corresponding to solve 
Eq.~(\ref{eq:sch-l}) for a spherically outgoing Coulomb wave. 
For complex momenta $ p=
p_R + {\rm i} p_I $ the energy also becomes complex $ E = Q - i\Gamma
/2 $. The boundary condition, Eq.~(\ref{eq:phase}) provides a
correlation between $\Gamma$ and $Q$ through the TPE potential. We get
$ \Gamma_{\rm pole} = 3.4(2) \, {\rm eV} $ for the S-matrix pole
width, fairly independently of the cut-off radius for $r_c \ll r_{\rm
min} \sim 3 {\rm fm}$.

Furthermore, for $p \ll 2 m_\pi $ we have the effective range
expansion~\footnote{Recent work~\cite{Higa:2008dn} provides a
one-to-one mapping from a Lagrangean with alpha and di-alpha fields to
Eq.~(\ref{eq:ere}) but disregarding pion exchanges.}
\begin{eqnarray}
\frac{ 2 \pi \cot \delta_0 (p)}{a_B(e^{2\pi \eta}-1)} + \frac2{a_B} h(\eta) = -
\frac1{\alpha_0} + \frac12 r_0 p^2 + \dots
\label{eq:ere}
\end{eqnarray} 
with $ h(x) $ the Landau-Smorodinsky
function~\cite{Rasche:1967,Kermode:1967} where the scattering length
$\alpha_0$ and the effective range $r_0$ are defined. From the
universal low energy theorem of Ref.~\cite{Valderrama:2005wv} in the
Coulomb case we obtain (in fm)
\begin{eqnarray}
r_0 = 1.03(1) - \frac{5.3(3)}{\alpha_0} + \frac{29(4)}{\alpha_0^2} \, .  
\end{eqnarray} 
The numerical coefficients depend on the TPE potential,
Eq.~(\ref{eq:tpe-pot}), plus Coulomb potential only. Using
Eq.~(\ref{eq:log-corr}) we find $\alpha_0^{\rm th} = -1210(70) {\rm
fm} $ and $r_0^{\rm th} = 1.03(1) {\rm fm}$, in reasonable agreement
with $\alpha_0= -1630(150) {\rm fm} $ and $r_0 =1.08(1) {\rm fm}$ from
a low energy analysis of the data~\cite{Rasche:1967}. The correlation
between $(Q,\Gamma)$ and $(\alpha_0,r_0)$ was pointed out long
ago~\cite{Kermode:1967,Kermode:1969} but has no predictive power. The
underlying chiral TPE potential correlates, in addition, $r_0$ with
$\alpha_0$ and $\Gamma$ with $Q$. At higher energies, however, further
ingredients are needed since for $E_{\rm LAB} = 3 {\rm MeV}$, we get
$\delta_0 = 141 \pm 2^0$ whereas $\delta_0=128.4 \pm 1^0$ from
data~\cite{RevModPhys.41.247}. The wavelength $1/p \sim 2.5 {\rm fm}$
corresponds to the two $\alpha$'s overlapping.


{\bf 7.} The leading nuclear structure contribution is given
by the lowest pionic excited $\alpha^*$-state which, however, lies
above the $t+p$ and $^3$He+n continuum ($M_{\alpha^*}-M_{\alpha}=25.28
{\rm MeV} $ whereas $M_t + M_p - M_\alpha = S_p = 19.8 {\rm MeV} $ and
$M_{^3He} + M_n - M_\alpha = S_n = 20.5 {\rm MeV} $). So, at long
distances the continuum states will dominate and thus we consider the
triangle diagrams of Fig.~\ref{fig:tpe} which involve three fermion
propagators. Taking the heavy particle limit $M_N, M_{3N}, M_\alpha
\to \infty$ with fixed separation energy $S_{N} = M_{3 N}+M_N-M_\alpha
$ the calculation reduces to the anomalous threshold contribution. If
we further take $g_{\alpha \alpha nn}=g_{\alpha \alpha pp} = g_4$ and
also $g_{\alpha p t }=g_{\alpha p ^3 He}= g_3 $ we get
\begin{eqnarray}
V_{\rm nucl.}(r) = - \frac{g^2}{4 \pi M_\alpha} \frac{e^{-2 \gamma r }}{r^2} 
\end{eqnarray} 
where $g^2 \sim g_3^2 g_4 $ (up to numerical factors) and the scale is
$ 1 /(2 \gamma) = 1/ 2 \sqrt{2 \mu_{N,3N} S_{N,3N}} \sim 0.58 {\rm
fm}$ and $S \sim 20 {\rm MeV}$ is the separation energy.  In fact,
this potential corresponds to the amplitude that one peripheral
nucleon of one $\alpha$-particle scatters with the other
$\alpha$-particle as a whole. Note the high level of degeneracy with
the long distance TPE potential, Eq.~(\ref{eq:v-long}) given the
similarity of both scales. A more complete study would require fixing
these couplings from nucleon-alpha scattering. Note that for $g^2 >
\pi $ this potential is also singular at short distances, and
similarly to the TPE case either $Q$ or $\alpha_0$, cannot be fixed
from the potential. We find that taking $g \sim 8$ the missing 1eV for
the width as well as the needed $0.04 {\rm fm }$ in the effective
range are obtained. Actually, a satisfactory description of
phase-shifts can be achieved (see Fig.~\ref{fig:phaseshift}).

In conclusion, the present analysis suggests that TPE as determined
from a chirally symmetric effective Lagrangean may indeed provide the
bulk of the $^8$Be lifetime, but competes with leading nuclear
structure effects at higher energies. A more precise statement
requires a better determination of the couplings which implies a
thourough study both of $\pi \alpha$ and $N\alpha$ scattering.


\bigskip
Supported by Spanish DGI and FEDER funds with grant FIS2005-00810,
Junta de Andaluc{\'\i}a grant FQM225-05, and EU Integrated
Infrastructure Initiative Hadron Physics Project contract
RII3-CT-2004-506078.


\bibliographystyle{aipproc}   


\begin{thebibliography}{22}
\expandafter\ifx\csname natexlab\endcsname\relax\def\natexlab#1{#1}\fi
\providecommand{\enquote}[1]{``#1''}
\expandafter\ifx\csname url\endcsname\relax
  \def\url#1{\texttt{#1}}\fi
\expandafter\ifx\csname urlprefix\endcsname\relax\def\urlprefix{URL }\fi
\providecommand{\eprint}[2][]{\url{#2}}

\bibitem[Afzal et~al.(1969)]{RevModPhys.41.247}
S.~A. Afzal, A.~A.~Z. Ahmad, and S.~Ali, \emph{Rev. Mod. Phys.} \textbf{41},
  247--273 (1969).

\bibitem[Tilley et~al.(2004)]{Tilley:2004}
D.~Tilley, et~al., \emph{Nucl. Phys.} \textbf{A745}, 155 (2004).

\bibitem[Frosch et~al.(1967)]{PhysRev.160.874}
R.~F. Frosch, J.~S. McCarthy, R.~E. Rand, and M.~R. Yearian, \emph{Phys. Rev.}
  \textbf{160}, 874--879 (1967).

\bibitem[Kermode(1965)]{Kermode:1965}
M.~W. Kermode, \emph{Nucl. Phys.} \textbf{A68}, 93--96 (1965).

\bibitem[Neudatchin et~al.(1971)]{Kukulin:1971}
V.~G. Neudatchin, V.~L. Kukulin, and V.~Korennoy, \emph{Phys. Lett.}
  \textbf{B34}, 581--584 (1971).

\bibitem[Friedrich(1981)]{Friedrich:1981ad}
H.~Friedrich, \emph{Phys. Rept.} \textbf{74}, 211--275 (1981).

\bibitem[Locher and Mizutani(1978)]{Locher:1978dk}
M.~P. Locher, and T.~Mizutani, \emph{Phys. Rept.} \textbf{46}, 43 (1978).

\bibitem[Arriola(2007)]{Arriola:2007de}
E.~Ruiz Arriola  (2007), \eprint{0709.4134}.

\bibitem[Weinberg(1979)]{Weinberg:1978kz}
S.~Weinberg, \emph{Physica} \textbf{A96}, 327 (1979).

\bibitem[Jenkins and Manohar(1991)]{Jenkins:1990jv}
E.~E. Jenkins, and A.~V. Manohar, \emph{Phys. Lett.} \textbf{B255}, 558--562
  (1991).

\bibitem[Ericson and Weise(1988)]{Ericson:1988gk}
T.~E.~O. Ericson, and W.~Weise, \emph{Pions and Nuclei}, Oxford, UK: Clarendon
  (1988), 1988.

\bibitem[Khankhasaev(1989)]{Khankhasaev:1989xf}
M.~K. Khankhasaev, \emph{Nucl. Phys.} \textbf{A505}, 717--754 (1989).

\bibitem[Shimodaya et~al.(1962)]{Shimodaya:1960}
I.~Shimodaya, R.~Tamagaki, and H.~Tanaka, \emph{Prog. Theor. Phys.}
  \textbf{27}, 793 (1962).

\bibitem[Fang-Landau and Locher(1973)]{FangLandau:1974wh}
S.~R. Fang-Landau, and M.~P. Locher, \emph{Nucl. Phys.} \textbf{B66}, 210--220
  (1973).

\bibitem[Ericson et~al.(1981)]{Ericson:1981sa}
M.~Ericson, P.~Guichon, and R.~D. Viollier, \emph{Nucl. Phys.} \textbf{A372},
  377 (1981).

\bibitem[Case(1950)]{Case:1950}
K.~M. Case, \emph{Phys. Rev.} \textbf{80}, 797--806 (1950).

\bibitem[Frank et~al.(1971)]{Frank:1971xx}
W.~Frank, D.~J. Land, and R.~M. Spector, \emph{Rev. Mod. Phys.} \textbf{43},
  36--98 (1971).

\bibitem[Pavon~Valderrama and Ruiz~Arriola(2006)]{Valderrama:2005wv}
M.~Pavon~Valderrama, and E.~Ruiz~Arriola, \emph{Phys. Rev.} \textbf{C74},
  054001 (2006), \eprint{nucl-th/0506047}.

\bibitem[Rasche(1967)]{Rasche:1967}
G.~Rasche, \emph{Nucl. Phys.} \textbf{A94}, 301 (1967).

\bibitem[Kermode(1967)]{Kermode:1967}
M.~W. Kermode, \emph{Nucl. Phys.} \textbf{A104}, 49--66 (1967).

\bibitem[Higa et~al.(2008)]{Higa:2008dn}
R.~Higa, H.~W. Hammer, and U.~van Kolck  (2008), \eprint{0802.3426}.

\bibitem[Kermode(1969)]{Kermode:1969}
M.~W. Kermode, \emph{Nucl. Phys.} \textbf{A134}, 336--346 (1969).

\end{thebibliography}

\end{document}